\documentclass[a4paper,12pt]{rspublic}
\usepackage{graphicx}
\usepackage{dcolumn}
\usepackage{bm}
\usepackage{subfig}
\usepackage{caption}
\usepackage{amsmath}
\usepackage{leftidx}
\usepackage{appendix}
\newcommand{\ud}{\mathrm{d}}
\begin{document}
\title{Triangular buckling patterns of twisted inextensible strips}
\author{A.~P.~Korte, E.~L.~Starostin and G.~H.~M.~van der Heijden
}
\affiliation{Centre for Nonlinear Dynamics, University College London, London WC1E 6BT, UK\\
a.korte@ucl.ac.uk\\
e.starostin@ucl.ac.uk\\
g.heijden@ucl.ac.uk
}
\date{\today}
\label{firstpage}
\maketitle
\begin{abstract}{twisted inextensible strip, triangular buckling pattern,
calculus of variations, boundary-value problem, load-displacement behaviour}
When twisting a strip of paper or acetate under high longitudinal tension,
one observes, at some critical load, a buckling of the strip into a regular
triangular pattern. Very similar triangular facets have recently been
observed in solutions to a new set of geometrically-exact equations describing
the equilibrium shape of thin inextensible elastic strips. Here we formulate
a modified boundary-value problem for these equations and construct
post-buckling solutions in good agreement with the observed pattern in
twisted strips. We also study the force-extension and moment-twist behaviour
of these strips by varying the mode number $n$ of triangular facets.
\end{abstract}

\section{Introduction}
\begin{figure}
\centering
\subfloat[][]{\label{fig:model_stripa}\includegraphics[width=0.3\textwidth]{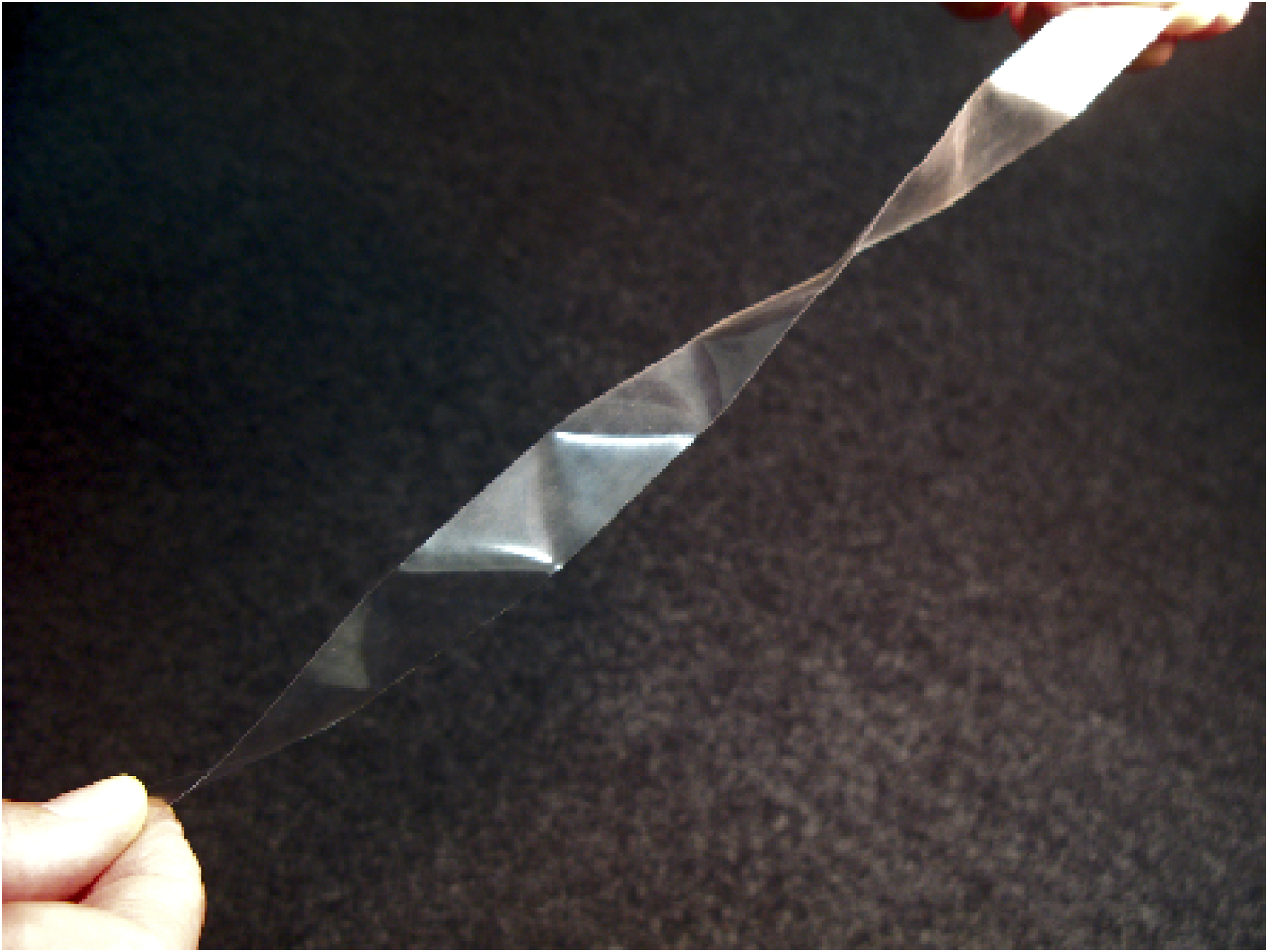}}
\subfloat[][]{\label{fig:model_stripb}\includegraphics[width=0.3\textwidth]{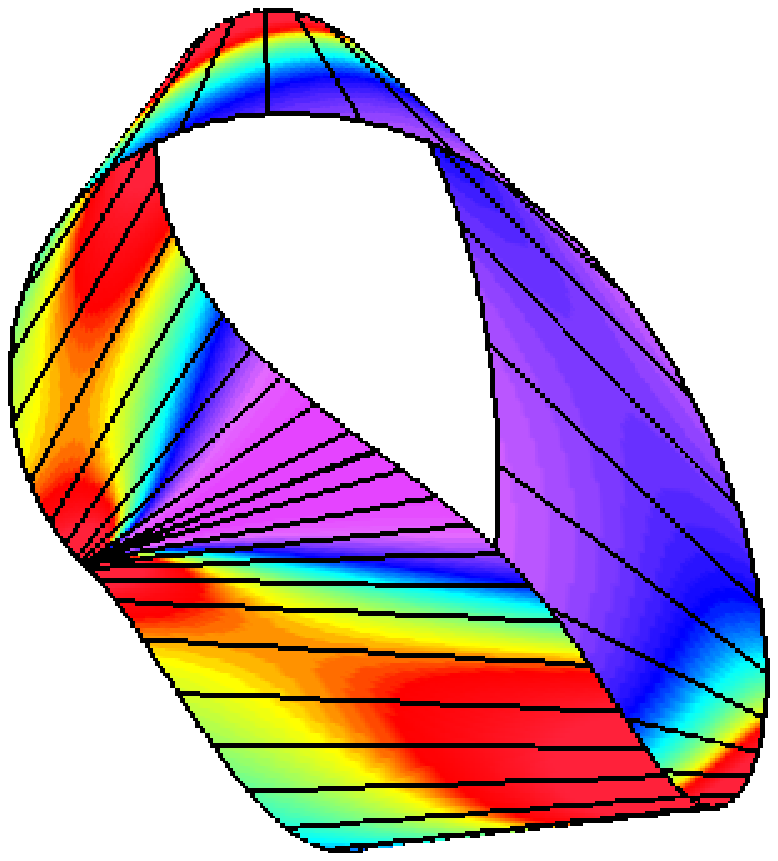}}
\subfloat[][]{\label{fig:model_stripc}\includegraphics[width=0.3\textwidth]{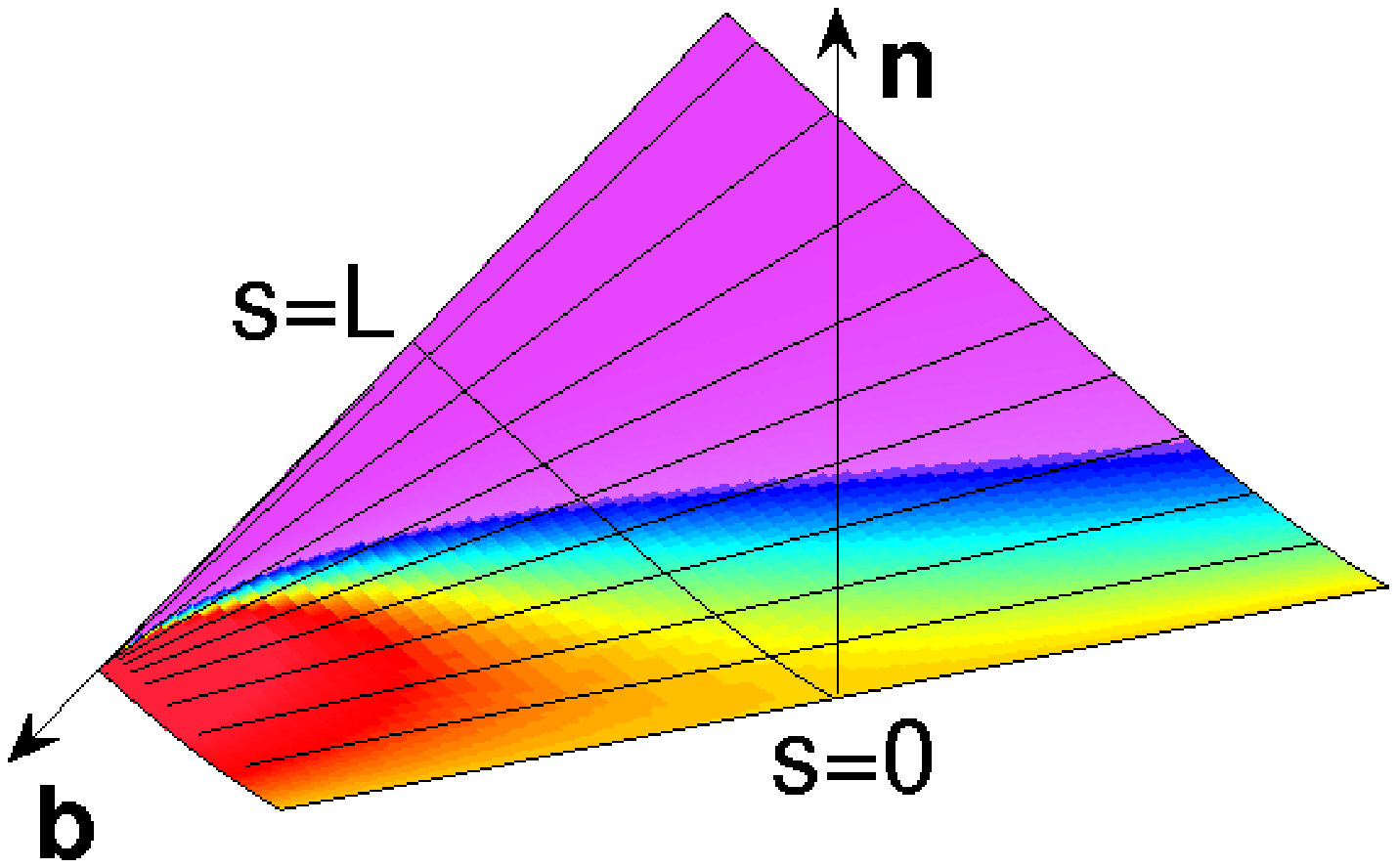}}
\caption{
(a) Twisted acetate model strip under tension.
(b) M\"obius developable structure of aspect ratio $2\pi$ with generators shown.
(c) Trapezoidal segment taken from the M\"obius structure to construct the periodic strip, with normal, binormal and centreline shown, $\eta'(0)=0, \eta(L)=0$.
The colouring changes according to the local bending energy density, from
violet for regions of low bending to red for regions of high bending. Note
the singularity on the edge of the strip.
}
\label{fig:model_strip}
\end{figure}

When twisting a strip of paper or acetate under high longitudinal tension,
one observes, at a critical load, a buckling of the strip into a regular
triangular pattern (see figure \ref{fig:model_strip}(a)). The deformation is
reversible. Sheets of paper or acetate are for practical purposes
inextensible and the observed pattern, consisting of helically stacked
nearly-flat triangular facets, appears to be nature's way of achieving global
twisting by means of local bending and minimal stretch. This mode of buckling
does not appear to have been reported in the literature. Perhaps this is
because the phenomenon occurs at relatively large twisting angles and under
relatively high tension (in order to suppress the more common looping
instability). The buckling pattern observed has ridges running at roughly
$45^{\circ}$ angles to the centreline of the strip. The ridges radiate out
from vertices on the edge of the strip where stress concentration occurs.
There is superficial similarity with the well-known Yoshimura or diamond
buckling pattern of thin cylindrical shells (Yoshimura 1930) but this pattern
requires compressive rather than tensile loading.

The phenomenon of buckling of an {\it extensible} elastic strip under tension
and twisting moment has been studied before. In the 1930s Green considered
buckling of twisted strips under constant tensile force, treating both the
case of zero and non-zero end force (Green 1936, Green 1937).
Buckling of a twisted orthotropic plate into a sinusoidal buckling pattern in
the longitudinal direction was studied in Crispino \& Benson (1986).
A numerical investigation of wrinkling of twisted plates was carried in
Mockensturm (2001), considering both the case of constant end-to-end distance
and the case of constant end force, and finding different results in the two
cases. In the latter case it was found that buckling may occur in both the
lateral and longitudinal directions.

A perturbation method was used recently to further explore the wrinkling
instability under small twist (Coman \& Bassom 2008). As the twist increases
from zero, the surface of a strip first becomes helicoidal, which causes
extensional forces near the edges while the core domain is in compression.
When a critical twist and tension are reached, the core domain of the strip
buckles into an oscillatory pattern. The surface near the edges remains
helicoidal; no strain localisation occurs.

The problem of bending and twisting of an {\it inextensible} flat plate has
also been studied before. Such a plate deforms isometrically and its surface
is therefore developable (i.e., has zero Gaussian curvature), making the
analysis more geometrical. Sadowsky developed a large-deformation theory of
narrow elastic strips as early as in 1930 (Sadowsky 1931). Approximate
equations for wide strips were derived in the mid-1950s by Mansfield (see
Mansfield 1989). These equations predict the distribution of generators of
the developable surface while ignoring the actual three-dimensional geometry.
This work was followed up by Ashwell (1962), where localisation of stresses
at two diagonally opposite corners is found for a strip in its first buckling
mode. The actual shape of the strip was not computed.

The constraint of zero Gaussian curvature causes the buckling patterns of
inextensible strips to differ strongly from the oscillatory buckling
patterns of extensible strips. The helicoidal shape of the edges of the strip
found in Coman \& Bassom (2008) is not a solution for inextensible strips
not even for infinitesimally small twist. Rather we observe a sequence of
relatively flat triangular domains which are not restricted to the core
of the strip. The edges show a sequence of points with high curvature.

It is worth noting that both responses, the smooth sinusoidal one and the
localised one, can be observed on the same paper strip model depending on the
environmental conditions. When the humidity is low and the paper is dry it
behaves like an inextensible material. When it is slightly wet it becomes
noticeably stretchable. Note also that the solution for a slightly extensible
strip is well described by the inextensible model almost everywhere except
for small domains where the stress concentrates. In this paper we compute
geometrically-exact developable solutions of inextensible strips.

A geometrically-exact set of equilibrium equations for the large deformation
of thin inextensible plates of finite width has recently been derived
(Starostin \& van~der Heijden 2007). The equations are ordinary differential
equations and obtained by using the inextensibility constraint to reduce
stresses and strains to the centreline of the strip. They are much easier to
analyse than the usual partial differential equations of elastic plate theory.
The new set of equations was used to solve a classical problem in mechanics,
namely to find the equilibrium shape of a developable M\"obius strip
(Starostin \& van~der Heijden 2007). Numerical solutions revealed the
existence, for any aspect ratio of the strip, of a nearly flat triangular
region associated with the (unique) inflection point of the strip (see
figure \ref{fig:model_strip}(b)). The triangular facet of the M\"obius strip
solution clearly resembles the facets of the buckling pattern of the twisted
strip in figure \ref{fig:model_strip}(a) and in this paper we use the new
system of equations to construct post-buckling solutions in good agreement
with experiment. The central idea is to modify the boundary-value problem for
the M\"obius strip such as to `cut out' the triangular (more precisely,
trapezoidal) region (see figure \ref{fig:model_strip}(c)) and to use
symmetry to reflect and multiply the elementary triangular facet into a
periodic triangular pattern. This procedure avoids having to integrate
numerically through the bending energy singularity associated with the vertex
of the triangle on the edge of the strip, as found in
Starostin \& van~der Heijden (2007) and analysed further in Hornung (2009).

Having obtained a post-buckling strip solution with a certain number, say
$n$, of triangular facets we then study the strip's force-extension and
moment-twist behaviour for various mode numbers $n$. Neither gravity nor
other distributed forces acting on the strip are taken into account in the
present work.

The paper is organised as follows. In Section 2 we formulate the
boundary-value problem for the centreline-reduced equations for a developable
strip. We also use symmetry properties of the solution to construct triangular
buckling patterns for the strip by concatenating elementary facets. In
Section 3 we numerically solve the boundary-value problem and compute
load-displacement curves for the post-buckling strip solutions for various
mode numbers $n$. In Section 4 we discuss our results and draw some
conclusions.

\section{The boundary-value problem}
A sufficiently thin elastic surface will deform by bending only (Witten 2007)
and therefore deform isometrically. If such a surface is flat in its
unstressed state it will remain so under deformation and therefore have zero
Gaussian curvature (Graustein 1966). It is said to be developable.

Now consider a rectangular sheet or strip.
If $\mathbf{r}(s)$ is a parametrisation of the centreline of the strip, $s$
being arclength, then
\begin{equation}\label{parametr2}
\begin{split}
\mathbf{x}(s,t)=\mathbf{r}(s)+t\left[\mathbf{b}(s)+\eta(s)\mathbf{t}(s)\right],\\
\tau(s)=\eta(s)\kappa(s), \qquad s\in [0,l], ~t\in[-w,w]
\end{split}
\end{equation}
is a parametrisation of an embedded developable strip of length $l$ and width
$2w$ (Randrup \& R{\o}gen 1996). Here $\mathbf{t}$ and $\mathbf{b}$ are two
unit vectors of the Frenet frame $\{\mathbf{t},\mathbf{n},\mathbf{b}\}$ of
tangent, principal normal and binormal to the centreline, while $\kappa$ and
$\tau$ are, respectively, the curvature and torsion of the centreline, which
uniquely specify (up to Euclidean motions) the centreline of the strip
(Graustein 1966). By equation \eqref{parametr2}, the surface, in turn, is
completely determined by the centreline of the structure. The straight
lines $s=\const$ are the generators of the surface, which make an angle
$\arctan(1/\eta)$ with the positive tangent direction. Given this
parametrisation, the mean curvature $M$ can be easily calculated, e.g., by
using the coefficients of the first and second fundamental forms of the
surface, themselves calculated from partial derivatives of $\mathbf{x}$ with
respect to $s$ and $t$ (Graustein 1966). The result is
\begin{equation}\label{mean_curv}
M=-\frac{\kappa}{2}\frac{1+\eta^2}{1+t\eta'},
\end{equation}
where the prime denotes differentiation with respect to arclength $s$.

Introducing rectangular co-ordinates $(u_1,u_2)$ by developing the surface
into a rectangle,
\begin{equation}\label{u1_u2}
u_1=s+t\eta(s), \qquad u_2=t,
\end{equation}
the bending energy of a strip of thickness $2h$ can be written as the
following integral over the surface of the strip (Love 1927):
\begin{equation}
U=2D\iint M^2 \,\mathrm{d}u_1\, \mathrm{d}u_2,
\label{energy3}
\end{equation}
where $D = 2 E h^3/[3 (1-\nu^2)]$ is the flexural rigidity, $E$ is Young's
modulus and $\nu$ is Poisson's ratio.
When $M$ (given by \eqref{mean_curv}) is substituted into \eqref{energy3},
and the co-ordinates changed to $s, t$, the $t$ integration can be carried
out (Wunderlich 1962) giving
\begin{equation}
U=Dw\int_0^L g(\kappa,\eta,\eta') \,\mathrm{d}s,
\label{energy4}
\end{equation}
with
\begin{equation}
g(\kappa,\eta,\eta')=\kappa^2\left(1+\eta^2\right)^2\frac{1}{2w\eta'}\log\left(\frac{1+w\eta'}{1-w\eta'}\right).
\label{energy5}
\end{equation}
In the zero-width limit, $w\to 0$, this reduces to the Sadowsky result
$g=\kappa^2(1+\eta^2)^2$ (Sadowsky 1931).

Minimisation of this elastic energy functional is a one-dimensional
variational problem cast in a form that is invariant under Euclidean motions.
In Anderson (1989) Euler-Lagrange equations for such geometric variational
problems are derived in a general context via a splitting of the cotangent
bundle $T^*J^{\infty}$ of the infinite jet bundle $J^{\infty}$ of a fibred
manifold, which induces a bigrading of the differential forms on
$J^{\infty}$ known as the variational bicomplex (see also Kogan \& Olver 2003).
In more physical terms they can be written in the form of six balance
equations for the (normalised) components of the internal force,
$\bm{F}=(F_t, F_n, F_b )^T$, and moment, $\bm{M}=(M_t, M_n, M_b )^T$, in
the directions of the Frenet frame, and two scalar equations
(Starostin \& van~der Heijden 2007, Starostin \& van~der Heijden 2009):
\begin{eqnarray}
&& \hspace{-1cm} \bm{F}'+\bm{\omega}\times\bm{F}=\bm{0}, \qquad
\bm{M}'+\bm{\omega}\times\bm{M}+\bm{t}\times\bm{F}=\bm{0},
\label{eqs_vector_b} \\
&& \hspace{-1cm} \partial_\kappa g+\eta M_t+M_b=0, \qquad
\left(\partial_{\eta'} g\right)'-\partial_\eta g-\kappa M_t=0,
\label{eqs_scalar_b}
\end{eqnarray}
where $\bm{\omega}=\kappa(\eta,0,1)^T$ is the Darboux vector. The equations
\eqref{eqs_vector_b} are nothing but the vectorial fixed-frame force and
moment balance equations $\mathbf{F}'=\mathbf{0}$,
$\mathbf{M}'+\mathbf{r}'\times\mathbf{F}=\mathbf{0}$, written out in the
Frenet frame. (Here we adopt the convention that bold roman symbols are used
for vectors while bold italic symbols are used for triples of components of
these vectors in the Frenet frame.) It follows immediately that
$\bm{F}\cdot\bm{F}$ and $\bm{M}\cdot\bm{F}$ are first integrals of the
equations. Note that the first equation in \eqref{eqs_scalar_b} is algebraic
in the variables $(\kappa$, $\eta$, $\eta')$, while the second is a
second-order ordinary differential equation (ODE) in $\eta$.

It will be of interest to a more general audience how the same equations can
be obtained from first principles using standard variational methods,
extending to a function of $\kappa, \tau$, $\kappa'$ and $\tau'$ the examples
in Capovilla \textit{et al}. (2002) (which covers functional dependence up to
$\kappa, \tau$ only). This is done in the appendix, where it is shown that it
is straightforward to accommodate the additional functional dependence on
$\kappa', \tau'$ without computing any new variations, by simply reusing the
results already given in Capovilla \textit{et al}. (2002) and
Langer \& Perline (1991). The additional terms generated by the dependence of
$g$ on $\kappa'$ and $\tau'$ are easily managed. It is straightforward to
extend this method to a functional involving any number of derivatives of
$\kappa$ and $\tau$, and hence obtain the closed-form expressions given in
Starostin \& van~der Heijden (2009).

The shape of the strip's centreline is found by first differentiating the
first equation in \eqref{eqs_scalar_b} in order to turn it into a differential
equation and then numerically solving equations \eqref{eqs_vector_b} and
\eqref{eqs_scalar_b} as a boundary-value problem (BVP) in conjunction with
three Euler-angle equations describing the evolution of the Frenet frame
relative to a fixed frame, and the centreline equation
$\mathbf{r}'=\mathbf{t}$. Taking Love's convention for the Euler angles
$(\theta,\psi,\phi)$ (Love 1927), the derivatives of the angles are related
to the curvature and torsion as
\begin{equation}\label{thetadot}
\begin{aligned}
\theta'&=\kappa\cos\phi,\\
\psi'&=\kappa\sin\phi/\sin\theta,\\
\phi'&=-\kappa\sin\phi\cot\theta+\kappa\eta,
\end{aligned}
\end{equation}
while the centreline equation in component form gives
\begin{equation}
\begin{aligned}
x'&=\sin\theta\cos\psi,\\
y'&=\sin\theta\sin\psi,\\
z'&=\cos\theta.
\end{aligned}
\end{equation}
Equation \eqref{thetadot} can be written down directly from the Darboux
vector by noting that Love's convention is the usual $y$-convention
(van~der Heijden \& Thompson 2000) in classical mechanics with $\psi$ and
$\phi$ interchanged (Goldstein 1980).
Alternatively it can be obtained from the Frenet-Serret equations.
The convention is such that the polar singularity (here at $\theta=0$)
usually associated with Euler angles does not cause any problems for the
solutions we are interested in (the flat strip will have $\theta=\pi/2$).
Alternatively a 4-parameter quaternion representation can be used to avoid
the singularity at the expense of a norm condition.

Before we specify the boundary conditions let's have a closer look at the
buckling pattern in figure \ref{fig:model_strip}(a) and the M\"obius strip
solution in figure \ref{fig:model_strip}(b).
The M\"obius strip solution has special points where either $\eta$ or $\eta'$
is zero. Points where $\eta'=0$ are called cylindrical points as the surface
is locally a cylinder, i.e., the mean curvature in \eqref{mean_curv} is
constant along the local generator. In figure \ref{fig:model_strip}(b) this
corresponds to a generator of constant colour. Points where $\eta=0$, by
contrast, are called conical because the edge of regression, on which nearby
generators intersect each other, has a cusp. At these conical points the
generator is perpendicular to the centreline, as follows from
\eqref{parametr2}. Clearly there must be at least one point where $\eta'=0$
between any two points with $\eta=0$. In fact, the M{\"o}bius strip has three
cylindrical points and three conical points, one of the latter being special
because it corresponds to the only inflection point of the centreline (where
$\kappa=0$). At this point the binormal component of both the force and the
moment are zero. Furthermore it is found that at the inflection point,
$|\eta'|\to 1/w$ (i.e., the edge of regression reaches the edge of the strip)
and the bending energy density $g$ diverges, i.e., we have stress
concentration. As is seen in figure \ref{fig:model_strip}(b), coming
out of this singular point is a nearly flat (violet) triangular region.

Now, turning to figure \ref{fig:model_strip}(a) we observe that the buckling
pattern consists of points of high stress located alternatingly on both
edges of the strip, while locally cylindrical ridges bound flat triangular
(more precisely, trapezoidal) regions similar to those found in the M\"obius
strip solution. This suggests that we can describe the buckling pattern by
a solution of the equations built up of alternating copies of the trapezoidal
section between the inflection point (where $\eta=0$) and the nearest
cylindrical point (where $\eta'=0$). A cut-out of this section is shown in
figure \ref{fig:model_strip}(c). Note that both bounding generators are of
constant colour, illustrating that the section can be reflected about both
end generators.

\begin{figure}
\includegraphics[width=0.95\textwidth]{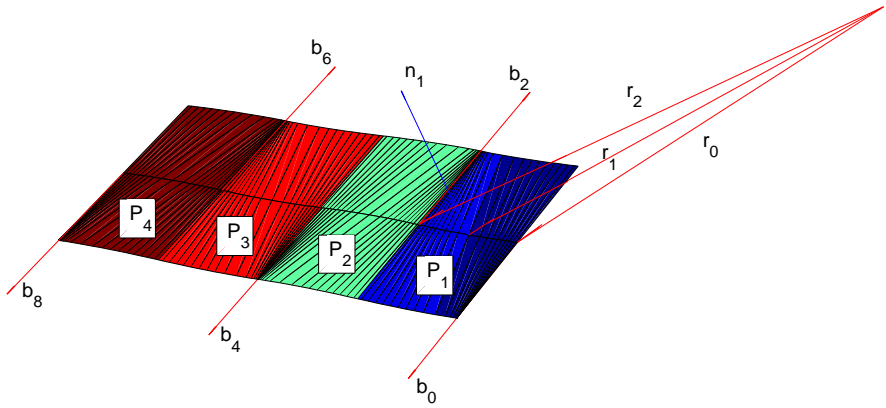}
\caption{\label{fig:rotations1} Mode $n=4$ symmetry operations. The first
half period is rotated about $\mathbf{n}_1$ to form the first period
$\mathrm{P}_1$. Rotating $\mathrm{P}_1$ about $\mathbf{b}_2$ produces the
second period $\mathrm{P}_2$. Rotating $\mathrm{P}_2$ about $\mathbf{b}_4$,
then $\mathrm{P}_3$ about $\mathbf{b}_6$, produces the final two periods of
the figure.}
\end{figure}

A two-step symmetry operation is therefore used to construct a strip of
length $2nL$ from a trapezoid of length $L$. Let the arclength parameter
of the centreline be $s=0$ at the cylindrical point ($\eta'=0$) and $s=L$
at the inflection point ($\eta=0$) (see figure \ref{fig:model_strip}(c)).
The first operation is a rotation through 180$^\circ$ of the trapezoid
about the normal $\mathbf{n}_1$ at $s=0$ (see also figure
\ref{fig:rotations1}). The original and the rotated trapezoid together make
a continuous surface, which forms one period of the full strip ($P_1$ of
figure \ref{fig:rotations1}), of length $2L$. The binormal to the strip at
the inflection point is denoted $\mathbf{b}_0$.

In the second step a symmetrical strip of $n+1$ periods is obtained from a
parent strip of $n$ periods by rotating an end period of the parent strip
180$^\circ$ about the end binormal. This binormal is located at the
inflection point of that end period and therefore aligned with the end
generator.
Thus strips of any period can be built up in this way by successively
reflecting an end period about its end binormal. The length of the centreline
of a symmetrical strip of $n$ periods made in this way is therefore $2nL$.
In particular, let $\mathbf{r}(0)\equiv\mathbf{r}_0$,
$\mathbf{r}(L)\equiv\mathbf{r}_1$ be respectively the position vectors of the
centreline at the inflection and cylindrical points of the trapezoid solution
of the BVP. Define a rotation of $\pi$ around the unit vector $\mathbf{g}$,
\begin{equation}
R_{\mathbf{g}}(\mathbf{a})=2\mathbf{g}(\mathbf{g}\cdot\mathbf{a})-\mathbf{a},
\end{equation}
then the first and subsequent periods are obtained by the iteration
\begin{equation}\label{refl_rules}
\begin{aligned}
\mathbf{r}_2=\mathbf{r}_1+R_{\mathbf{n}_1}(\mathbf{r}_0-\mathbf{r}_1),& \quad \mathbf{b}_2=R_{\mathbf{n}_1}(\mathbf{b}_0),\\
\mathbf{r}_{2i}=\mathbf{r}_{2i-2}+R_{\mathbf{b}_{2i-2}}(\mathbf{r}_{2i-4}-\mathbf{r}_{2i-2}),& \quad \mathbf{b}_{2i}=R_{\mathbf{b}_{2i-2}}(\mathbf{b}_{2i-4}),
\end{aligned}
\end{equation}
where $i=2,\ldots,n$. The $\mathbf{b}_{2i}$ are then the unit binormals at
the inflection points of the periods of the strip.
The reflection rules in \eqref{refl_rules} are for the centreline, but can be
easily generalised to the whole surface, as was done in
Starostin \& van~der Heijden (2007) to obtain closed-strip solutions from the
trapezoid solution of the BVP. These reflection rules are shown in
figure \ref{fig:rotations1} for the example $n=4$ with the relevant binormals
shown. The initial trapezoid (with one boundary aligned with $\mathbf{b}_0$)
is rotated about $\mathbf{n}_1$ giving the
first period $P_1$. This first period is then rotated about $\mathbf{b}_2$ to
give the second period $P_2$, which itself is rotated about $\mathbf{b}_4$ to
give the third period $P_3$. Finally the third period is rotated about
$\mathbf{b}_6$ to give the fourth period $P_4$.

It remains to formulate the boundary conditions for the initial trapezoid.
At $s=0$ these are
\begin{equation}
\begin{aligned}\label{bcs_seq0}
F_n(0)&=0,& \quad M_n(0)&=0,& \quad \eta'(0)&=0, \\
x(0)&=0,& \quad y(0)&=0,& \quad z(0)&=0,
\end{aligned}
\end{equation}
where the values for $x$, $y$ and $z$ fix an arbitrary position in space, and
\begin{equation}\label{bcs_seq0nl}
M_b(0)=-\eta(0) M_t(0)-2\kappa(0)(1+\eta^2(0))^2,
\end{equation}
which can be read off as the $s=0$ limit of the first equation of
\eqref{eqs_scalar_b}, where $\eta'(0)=0$. Equation \eqref{bcs_seq0nl} is
required to fix the integration constant for the ODE obtained by
differentiating the first equation of \eqref{eqs_scalar_b}.
(Note that by Taylor expanding the second of equations \eqref{eqs_scalar_b}
around $s=0$, one can also show that
$M_t(0)=\frac{2}{3}(1+\eta^2(0))\kappa(0)(-6\eta(0)+w^2(1+\eta^2(0))\eta''(0))$).

The boundary conditions at $s=L$ are
\begin{equation}\label{bcs_seqL}
\begin{aligned}
\kappa(L)&=0,& F_b(L)&=0,& M_b(L)&=0,&  \\
\theta(L)&=\pi/2,& \psi(L)&=0,& \phi(L)&=\pi,
\end{aligned}
\end{equation}
where the angles fix an arbitrary orientation of the strip in space.

The force and moment boundary conditions in \eqref{bcs_seq0} and
\eqref{bcs_seqL} are enforced by the rotational symmetry. Since
$\mathbf{n}_1$ and $\mathbf{b}_0$ are local axes of rotational (or reflection)
symmetry, any component of force and moment along those axes must vanish.
This guarantees continuity of forces and moments in the strip and therefore
yields a valid solution of the mechanical problem.

The remaining boundary conditions for the system of ODEs are the projections
of the end force and moment of the strip on the end-to-end unit
vector $\hat{\mathbf{e}}=\mathbf{e}/|\mathbf{e}|$,
$\mathbf{e}=\mathbf{r}_{2n}-\mathbf{r}_0$,
\begin{equation}\label{bcs_F2M2}
\begin{aligned}
\mathbf{F}(L)\cdot\hat{\mathbf{e}}&=\bar{F},& \quad \mathbf{M}(L)\cdot\hat{\mathbf{e}}&=\bar{M},
\end{aligned}
\end{equation}
where $\bar{F}, \bar{M}$ are also continuation (control) parameters.
Equations \eqref{bcs_seq0}--\eqref{bcs_F2M2} give a final count of $15$
boundary conditions for the same number of first-order ODEs.

The BVP is solved by continuation of each of the boundary conditions
in \eqref{bcs_F2M2} in turn. To obtain the starting solution of the
continuation, the numerical solution of the M\"obius strip between the
inflection point and nearest cylindrical point is continued in $F_n(0)$
and $M_n(0)$ until equations \eqref{bcs_seq0} are satisfied.

There are significant numerical difficulties solving this BVP as both ends
of the integration interval have a singularity. The boundary condition
$\kappa(L)=0$ enforces an inflection point at $x=L$. At this point the torsion
$\tau$ must also be zero and in fact it must go to zero, in $s$, faster than
the curvature $\kappa$ (Randrup \& R{\o}gen 1996). Therefore we also have
$\eta(L)=0$. In practice, to compute a starting solution we first compute an
approximate solution with $\kappa(L)\simeq 0.1$ to stay away from the
singularity at $x=L$. When all other boundary conditions are satisfied we
`pull' the solution into the singularity by continuing $\kappa(L)$ to zero
as far as possible, typically reaching values of 0.001. At this point we
typically have $\eta(L)\simeq 0.002$, while $|\eta'(L)|-1/w$, the distance
from the singularity, is typically as small as $10^{-6}$.

At the other singularity, at $s=0$, numerical convergence requires Taylor
expansions (up to fourth order in our case) of the left-hand sides of
equations \eqref{eqs_scalar_b} about $\eta'=0$ to be used for a small interval
around $s=0$. In addition, to improve convergence, the absolute value of the
logarithm is taken in equation \eqref{energy5}. We note that it has not
proved possible (for instance by similar Taylor expansions) to remove the
singularity in \eqref{eqs_scalar_b} at $\eta'=1/w$.

As a check on the numerical results, the first integrals $\bm{F}\cdot\bm{F}$
and $\bm{M}\cdot\bm{F}$ are typically found to be constant to within
$10^{-9}$.

Once the BVP has been solved, the surface of the strip is obtained from
\eqref{parametr2} and the reflection rules \eqref{refl_rules}. We note that
no measures in the above formulation are taken to prevent self-intersection
of the strip.

\section{Numerical results: response of the strip to applied loads}
The AUTO continuation software (Doedel \textit{et al}. 2007) was used to
compute response curves of applied force
$\bar{F}\equiv\mathbf{F}(L)\cdot\hat{\mathbf{e}}$ against end-to-end distance
of the strip $|\mathbf{e}|$, and applied moment
$\bar{M}\equiv\mathbf{M}(L)\cdot\hat{\mathbf{e}}$ against the accumulated
twist angle $\alpha$. As the component of $\mathbf{b}_0$ perpendicular to
$\mathbf{e}$ is
$\mathbf{a}_0\equiv\mathbf{b}_0-(\mathbf{b}_0\cdot\hat{\mathbf{e}})\hat{\mathbf{e}}$,
with $\mathbf{a}_{2n}$ similarly defined, the twist angle $\alpha$ is the
angle between the components of the two end-of-strip binormals perpendicular
to the end-to-end vector of the strip, and therefore
$\cos\alpha=\mathbf{a}_0\cdot\mathbf{a}_{2n}/(|\mathbf{a}_0||\mathbf{a}_{2n}|)$.

\begin{figure}
\centering
$
\begin{array}{c}
\subfloat[][]{\includegraphics[width=0.95\textwidth]{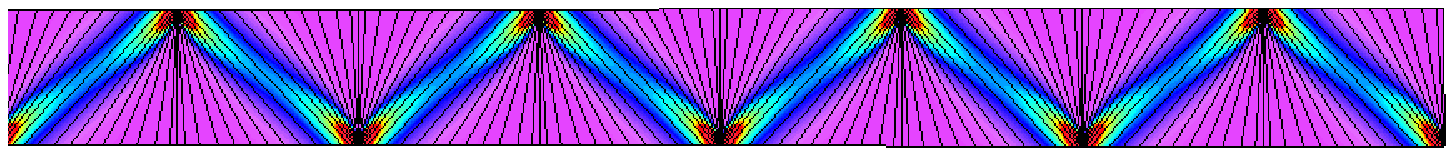}}\\
\subfloat[][]{\includegraphics[width=0.95\textwidth]{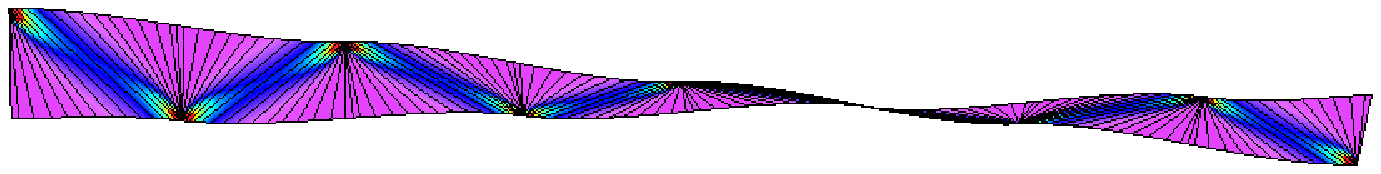}}\\
\subfloat[][]{\includegraphics[width=0.95\textwidth]{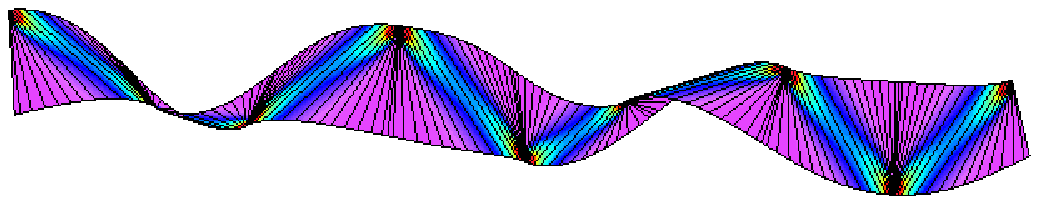}}
\end{array}$
\caption{Shapes of strips for $n=8$. 
(a) $\bar{F}=13.42$, $\bar{M}=0.162$.
(b) $\bar{F}=6.80$, $\bar{M}=2.71$.
(c) $\bar{F}=0.197$, $\bar{M}=12.15$.
}
\label{fig:Surface_neq_8}
\end{figure}

Under a distance rescaling $s\to s'=sk$, the force and moment scale as
$\mathbf{F}\to\mathbf{F}/k^2$ and $\mathbf{M}\to\mathbf{M}/k$ ($\mathbf{M}$
scales as $\kappa$, cf. equation \eqref{bcs_seq0nl}), $\kappa$ scales as
$\kappa\to\kappa/k$, whereas $\eta$ and the Euler angles are scale invariant.
Rescaling can be used to obtain strip solutions of a given period from
solutions of another different period, with the same aspect ratio, via a $w$
continuation. For example, starting from a strip of length $2nL$, width $2w$,
and period $n=2m$, selecting $n/2$ consecutive periods gives a strip of
length $nL$, continuing this solution to $w$ and rescaling $s\to 2s$ gives a
period $n/2$ strip with the same aspect ratio as the original strip. All
strips are scaled to an aspect ratio of $2nL/2w=10.5300$ unless otherwise
stated.

Many different strip shapes are obtained depending on the boundary conditions
\eqref{bcs_F2M2}. In figure~\ref{fig:Surface_neq_8} are shown three shapes
of a strip with $n=8$, one under a high axial tension, one under a relatively
high axial moment and an intermediate one which is in reasonably good
agreement with the experiment in figure~\ref{fig:model_strip}(a). Here, and
in all surface plots in this paper, the colouring varies according to the
bending energy density, from violet for regions of low bending to red for
regions of high bending (scales are individually adjusted).

\begin{figure}
\centering
$
\begin{array}{cc}
\subfloat[][]{\includegraphics[scale=0.4]{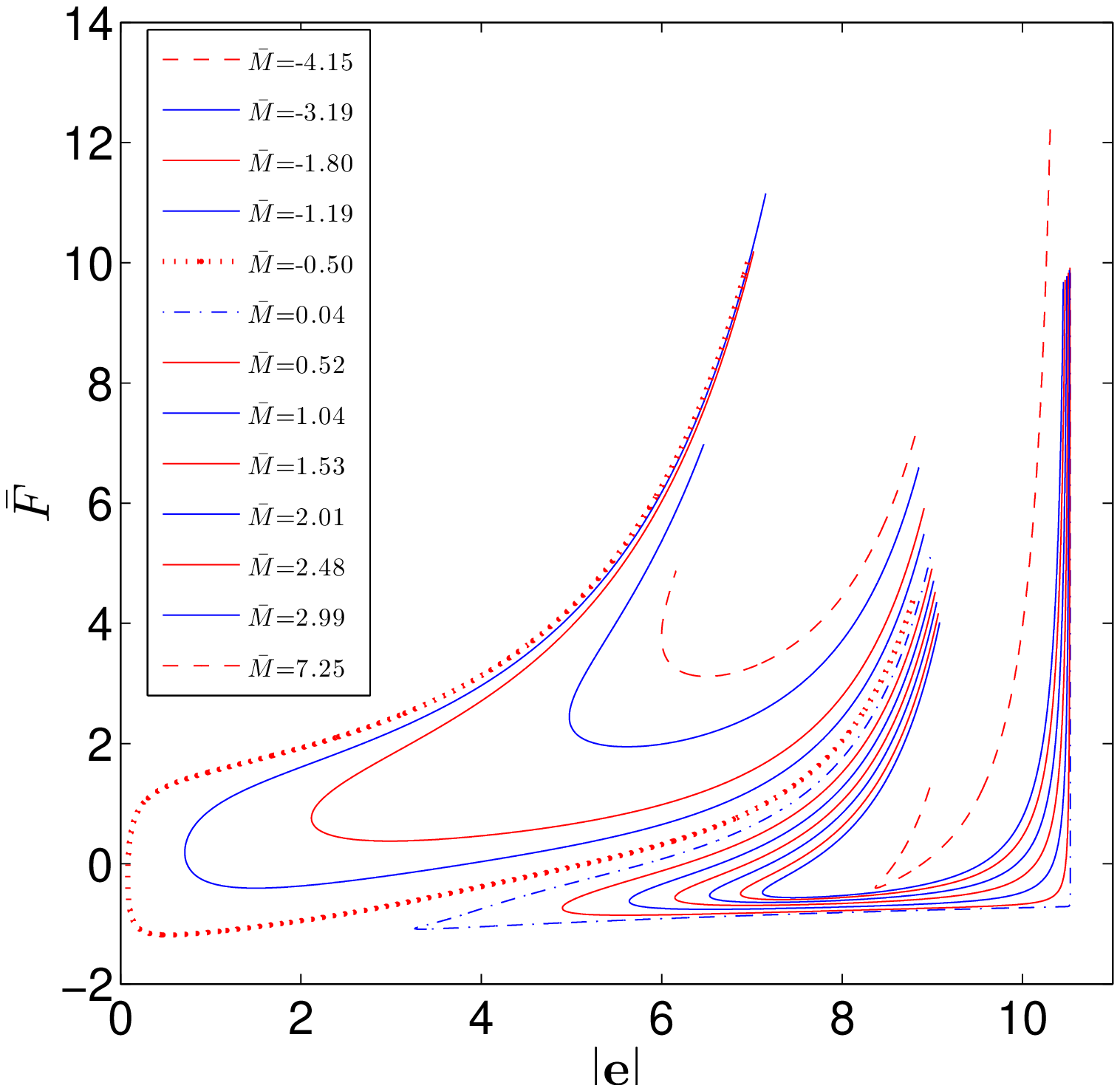}} &
\subfloat[][]{\includegraphics[scale=0.4]{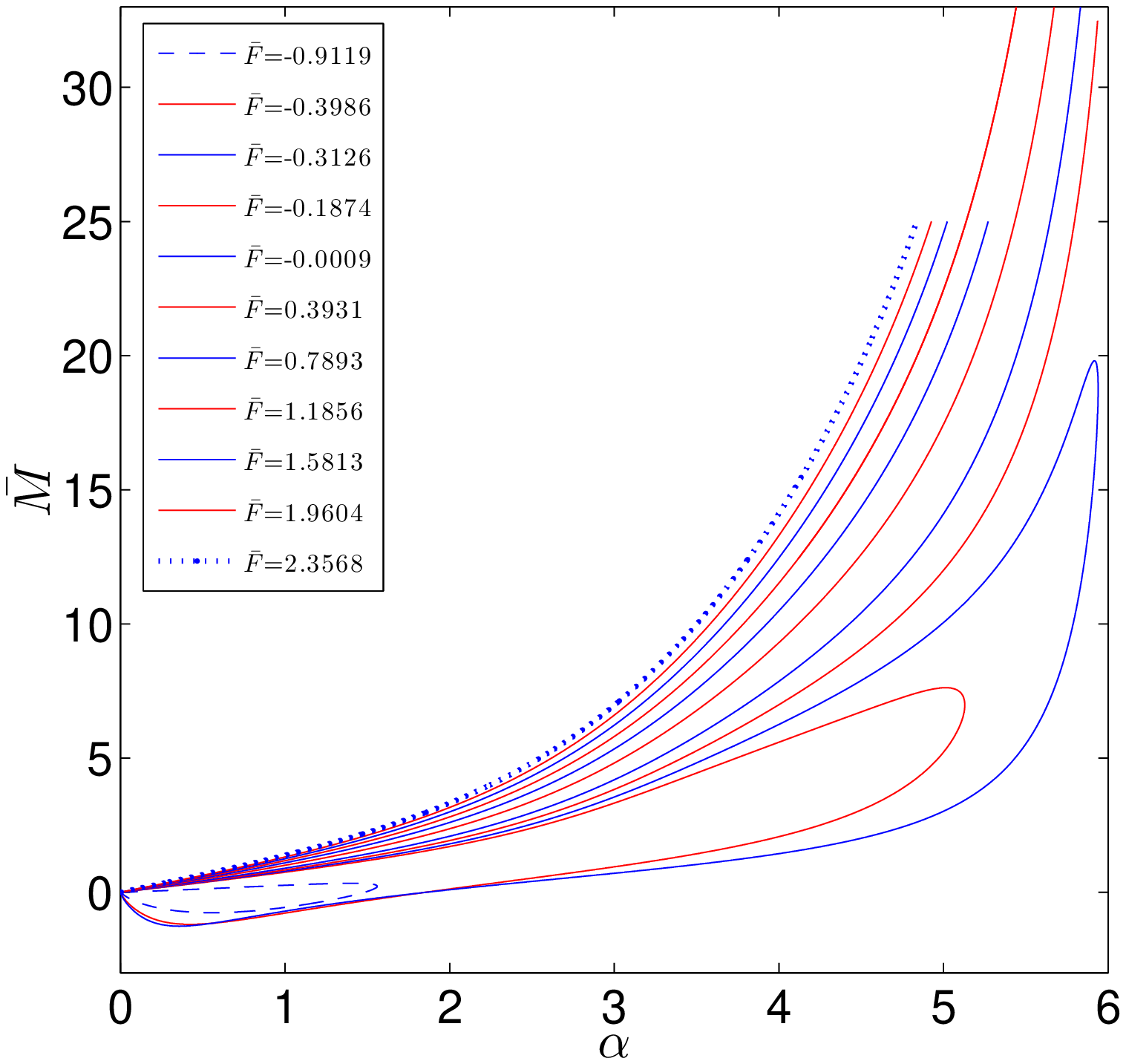}}
\end{array}$
\caption{$n=2$ mode.
(a) Force $\bar{F}$ versus scaled end-to-end distance, $|\mathbf{e}|$.
The curves of constant $\bar{M}$ form a nested sequence on the plot, with $\bar{M}$ increasing from the top middle dashed curve ($\bar{M}=-4.15$) to the dotted curve, then from the dot-dashed curve to the dashed curve on the right.
(b) Moment $\bar{M}$ versus twist angle $\alpha$ (in radians).
The curves of constant $\bar{F}$ form a nested sequence on the plot, with $\bar{F}$ increasing from the dashed curve ($\bar{F}=-0.9119$) to the dotted curve.
}
\label{fig:F2vsde2e_M2vsalpha_neq2}
\end{figure}

\begin{figure}
\centering
$
\begin{array}{cc}
\subfloat[][]{\includegraphics[width=0.45\textwidth]{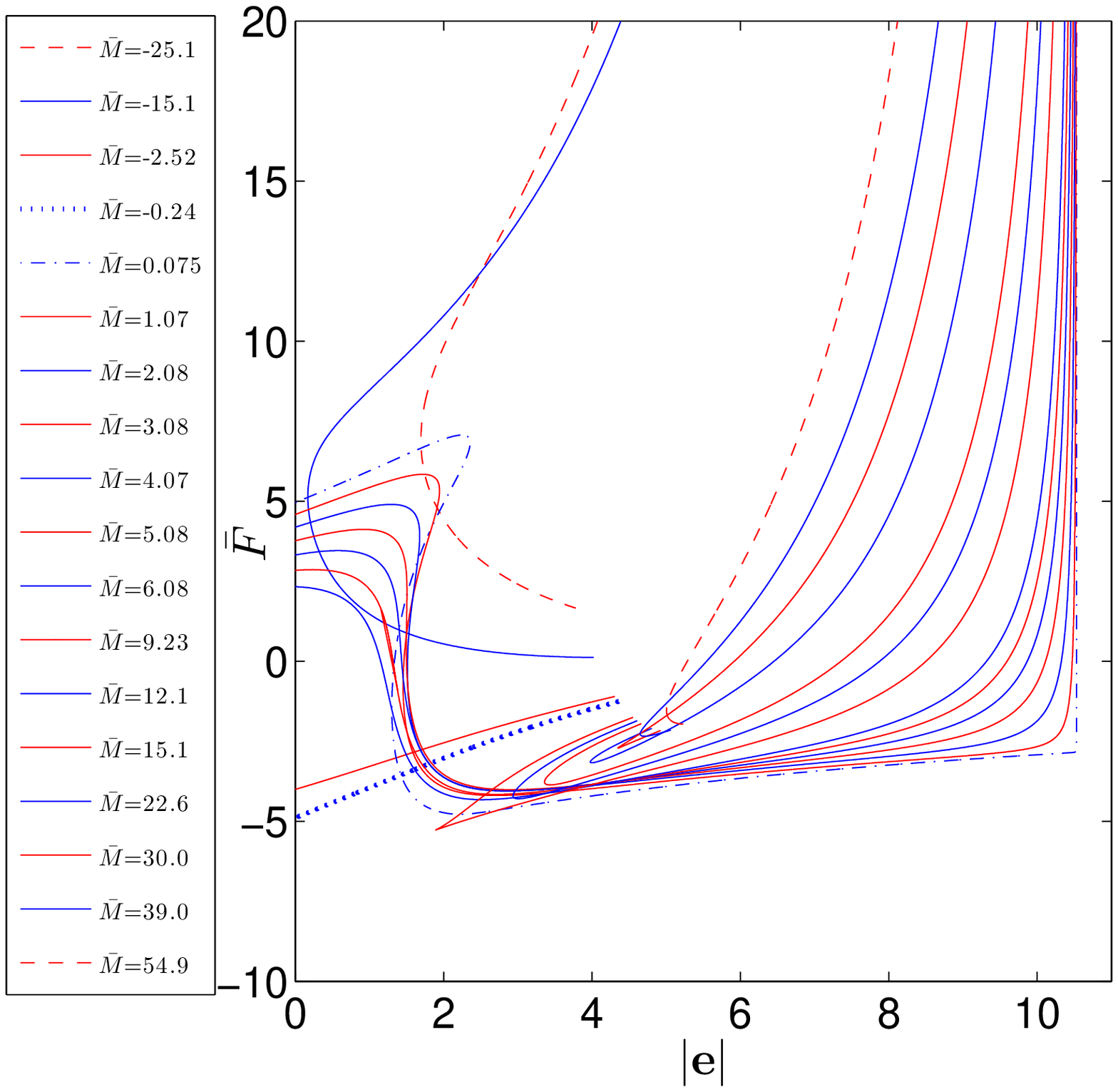}} &
\subfloat[][]{\includegraphics[width=0.45\textwidth]{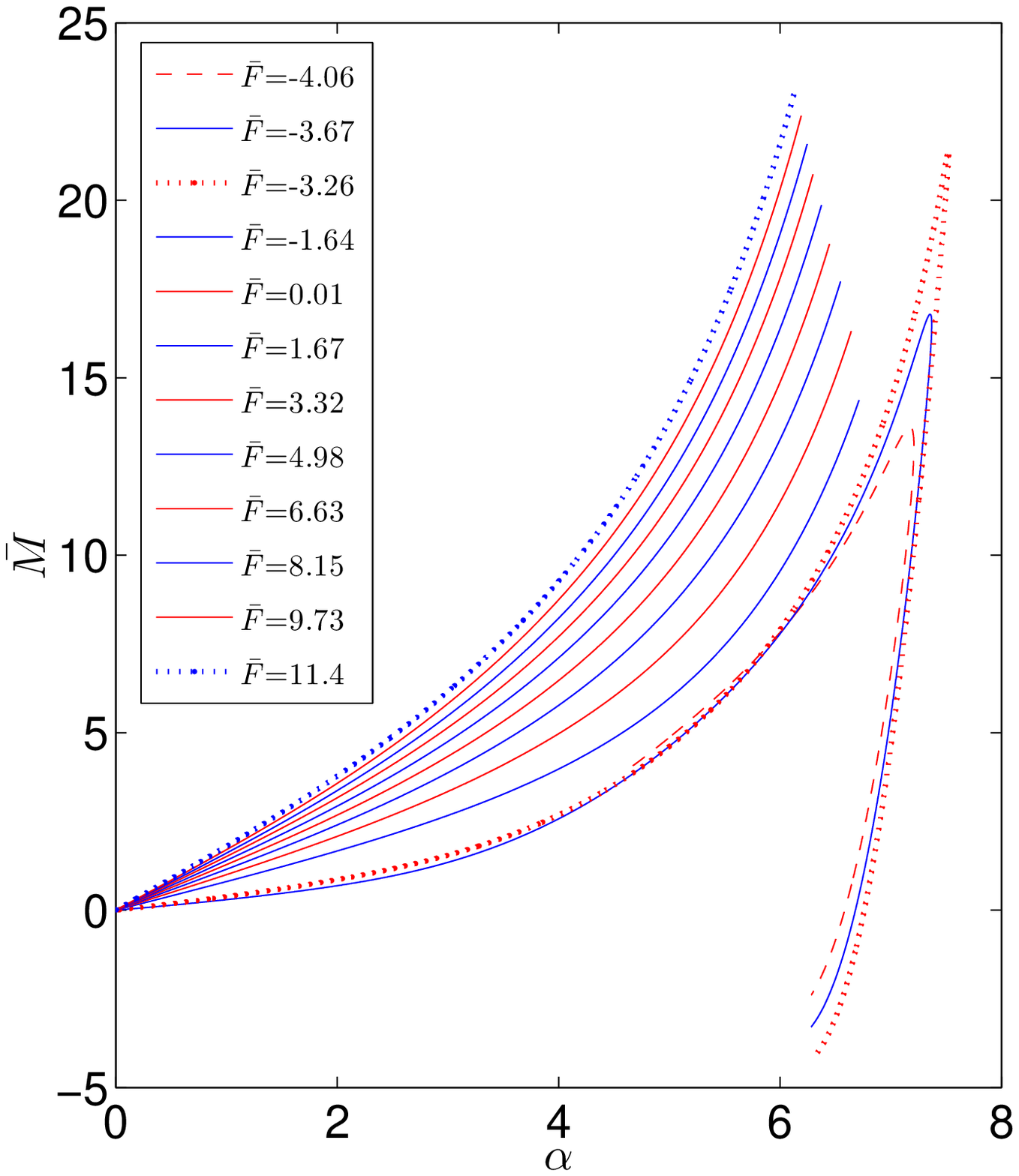}}
\end{array}$
\caption{$n=4$ mode.
(a) Force $\bar{F}$ versus scaled end-to-end distance, $|\mathbf{e}|$.
The curves of constant $\bar{M}$ form a sequence on the plot, with $\bar{M}$ increasing from the top left dashed curve ($\bar{M}=-25.1$) to the dotted curve, then from the dot-dashed curve to the top right dashed curve.
(b) Moment $\bar{M}$ versus twist angle $\alpha$ (in radians).
The curves of constant $\bar{F}$ form a sequence on the plot, with $\bar{F}$ increasing from the dashed curve ($\bar{F}=-4.06$) to the dotted curve ($\bar{F}=-3.26$), then to the top dotted curve.
}
\label{fig:F2vsde2e_M2vsalpha_neq4}
\end{figure}

\begin{figure}
\centering
$
\begin{array}{cc}
\subfloat[][]{\includegraphics[width=0.45\textwidth]{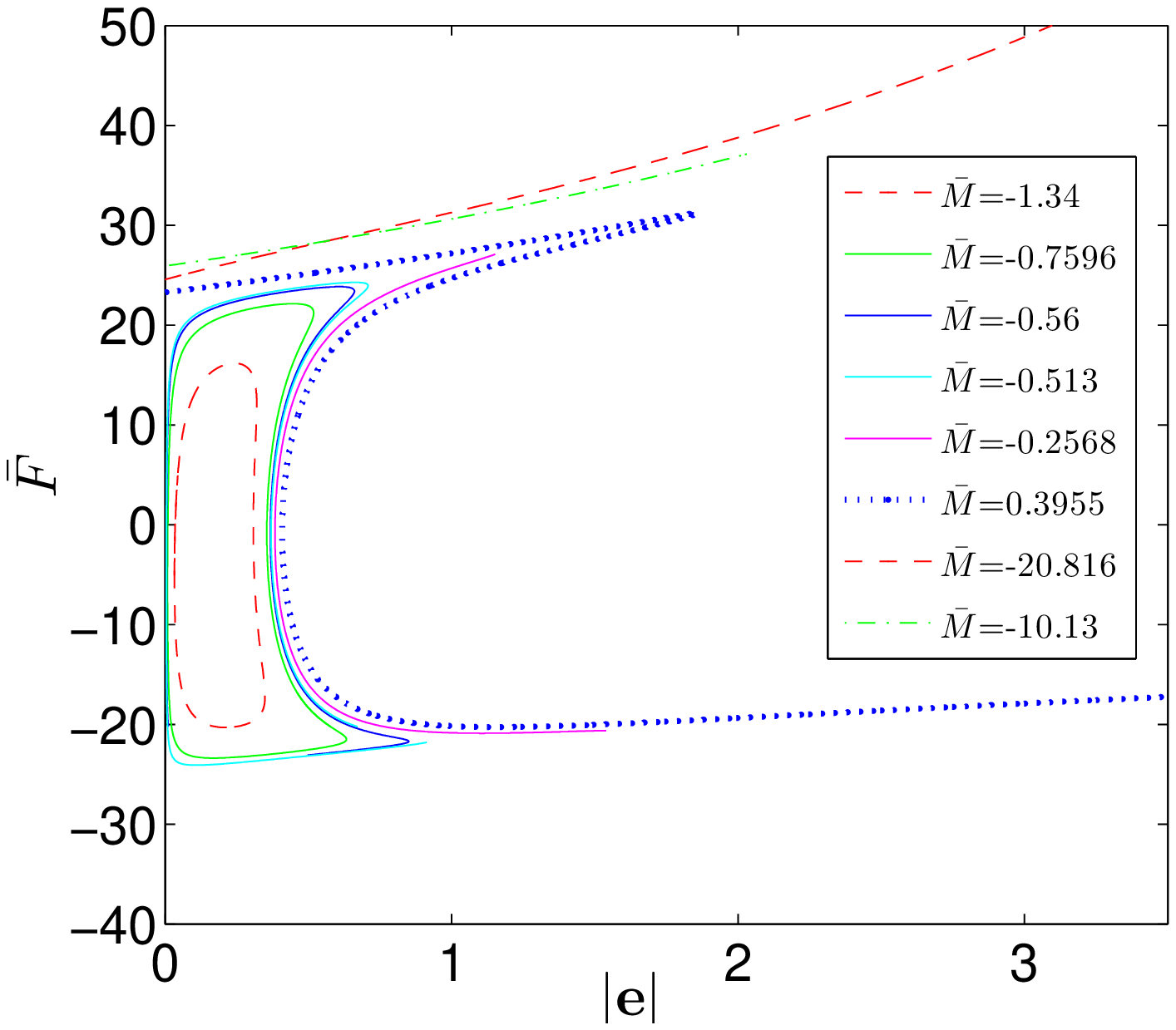}} &
\subfloat[][]{\includegraphics[width=0.45\textwidth]{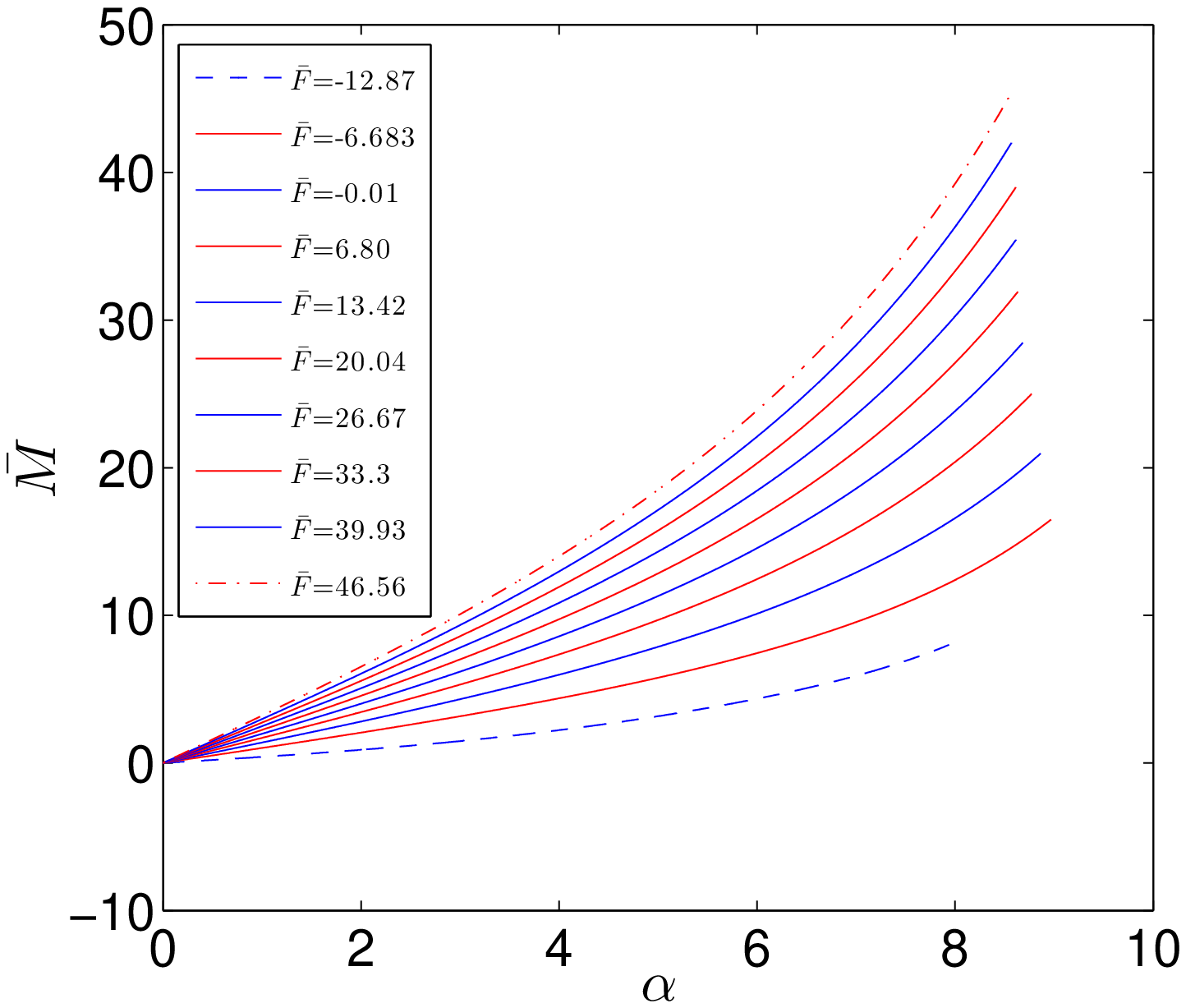}}
\end{array}$
\caption{$n=8$ mode.
(a) Force $\bar{F}$ versus scaled end-to-end distance, $|\mathbf{e}|$.
The curves of constant $\bar{M}$ form a nested sequence on the plot, with $\bar{M}$ increasing from the left dashed loop ($\bar{M}=-1.34$) to the dotted curve, then from the dot-dashed curve to the top dashed curve.
(b) Moment $\bar{M}$ versus twist angle $\alpha$ (in radians).
The curves of constant $\bar{F}$ form a nested sequence on the plot, with $\bar{F}$ increasing from the dashed curve ($\bar{F}=-12.87$) to the dot-dashed curve.
}
\label{fig:F2vsde2e_M2vsalpha_neq8}
\end{figure}

Continuation results for modes $n=2$, 4, 8 are shown in
figures~\ref{fig:F2vsde2e_M2vsalpha_neq2}--\ref{fig:F2vsde2e_M2vsalpha_neq8}
(for aspect ratio $2nL/2w=10.5300$, except for
figure~\ref{fig:F2vsde2e_M2vsalpha_neq8}(a), which has aspect ratio $42.30$).
On the left of each figure is shown the force response for the sequence of
axial moments ($\bar{M}$) shown on the legend. The length scale is arbitrary
and is such that $L=0.6581$.
On the right of each figure is shown the moment response for the sequence
of axial forces ($\bar{F}$) shown on the legend. These curves were obtained by
continuation, where $\bar{F}$ was varied, keeping $\bar{M}$ fixed or vice
versa. Exceptions to this are some curves in
figure~\ref{fig:F2vsde2e_M2vsalpha_neq4}(a,b) and
figure \ref{fig:F2vsde2e_M2vsalpha_neq8}(a), where both $\bar{F}$ and
$\bar{M}$ were allowed to vary, only so that a continuation could be started
at an arbitrary point on the graph (at which point one of $\bar{F},\bar{M}$
would then be held fixed). This accounts for the curve-crossing in this
figure.

\begin{figure}
\centering
\includegraphics[width=0.99\textwidth]{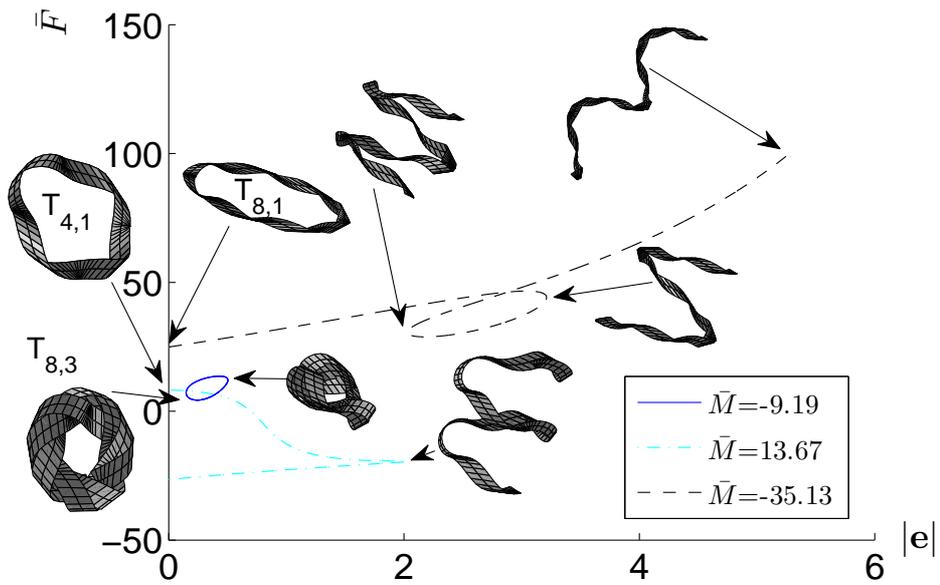}
\caption{$n=8$ mode.
Force $\bar{F}$ versus scaled end-to-end distance, $|\mathbf{e}|$, showing
strip solutions.
}
\label{fig:F2vsde2e_M2vsalpha_neq8_2}
\end{figure}

\begin{figure}
\subfloat[][]{\includegraphics[width=0.45\textwidth]{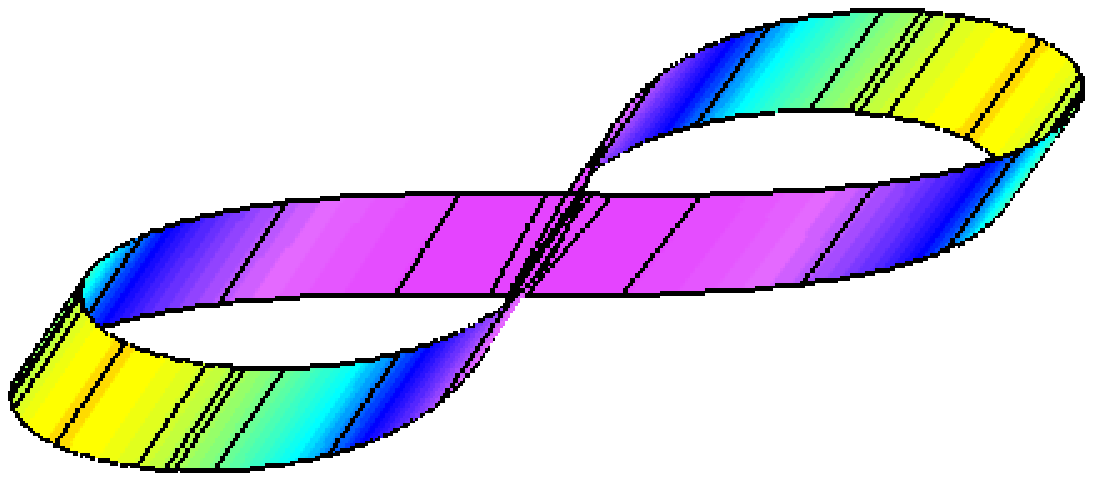}}
\subfloat[][]{\includegraphics[width=0.45\textwidth]{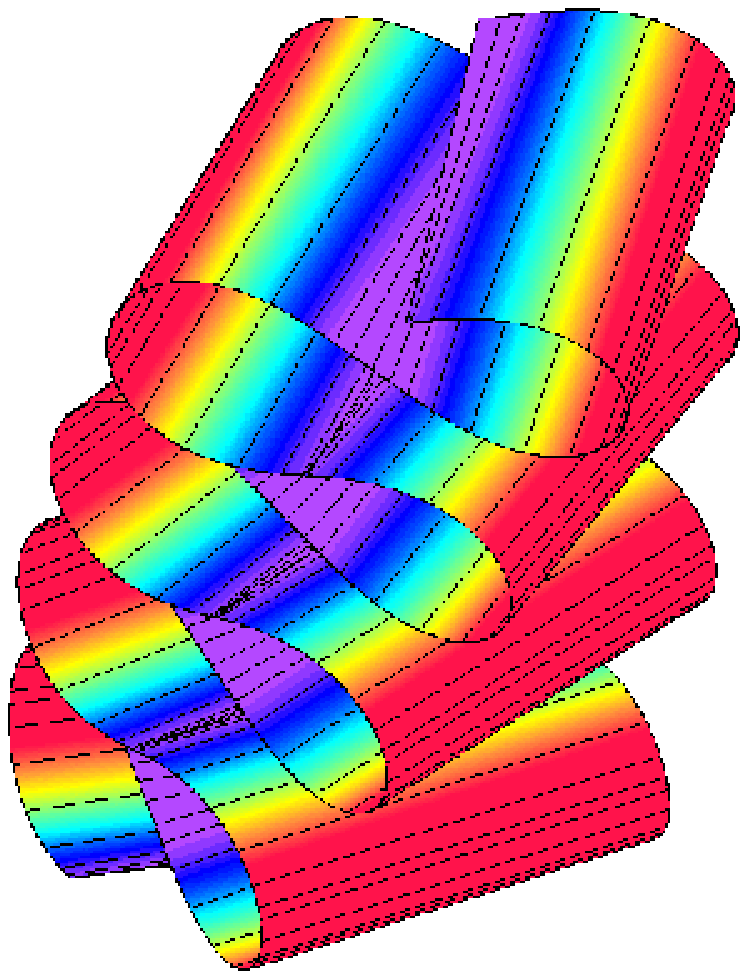}}
\caption{Shapes of strips.
(a) $n=2$, figure-of-eight.
(b) $n=8$, $\bar{F}=-19.975$, $\bar{M}=0.1424$.
}
\label{fig:Surface_neq_4}
\end{figure}

In figure~\ref{fig:F2vsde2e_M2vsalpha_neq2}(a), for the $n=2$ mode, the
sequence of curves at constant negative applied end moment is the group at
the bottom left, with the moment increasing from top (large negative) to
bottom (near zero). The group of curves at the bottom right of the figure
is for constant positive applied end moment, with the moment increasing from
bottom (near zero) to top (large positive). In
figure~\ref{fig:F2vsde2e_M2vsalpha_neq2}(b) is a sequence of curves at
constant applied end force, increasing from bottom (negative) to top
(positive). The curves for smallest (negative) constant force start at zero
moment, have a maximum twisting angle, which returns to zero twist angle,
which is the figure-of-eight shown in figure~\ref{fig:Surface_neq_4}(a).

In figure~\ref{fig:F2vsde2e_M2vsalpha_neq4}(a), for the $n=4$ mode, the
sequence of curves at constant negative applied end moment is the group at
the bottom left, with the moment increasing from top (large negative) to
bottom (near zero). The group of curves at the bottom right of the figure
is for constant positive applied end moment, with the moment increasing
from bottom (near zero) to top (large positive).
In figure~\ref{fig:F2vsde2e_M2vsalpha_neq4}(b) is a sequence of curves at
constant applied end force, increasing from bottom (negative) to top
(positive), showing the response curve of applied end moment to the twist
angle $\alpha$. The three lowest negative applied force continuations
terminate at $\alpha=2\pi$, again forming a closed strip, this time the
$T_{4,1}$ elastic torus ribbon knot.

In figure~\ref{fig:F2vsde2e_M2vsalpha_neq8}(a), for the $n=8$ mode, is a
sequence of curves at constant applied end moment (positive and negative).
This picture is more complicated than the other modes. Some of the curves
terminate on the vertical axis, i.e., when the end-to-end distance vanishes.
These closed ribbon solutions are invariably torus ribbon knots. For example,
the upper dashed curve ($\bar{M}=-20.816$) terminates in the double cover of
the $T_{4,1}$ elastic torus ribbon knot. The series of nested closed curves
have a quadruple cover of the figure-of-eight solution where the curves
approach vanishing end-to-end distance.
In figure~\ref{fig:F2vsde2e_M2vsalpha_neq8}(b) is a sequence of curves at
constant applied end force, increasing from bottom (negative) to top
(positive), showing the response curve of applied end moment to the twist
angle $\alpha$.

In figure~\ref{fig:F2vsde2e_M2vsalpha_neq8_2}, for the $n=8$ mode, is a
sequence of curves at constant applied end moment, with some of the resulting
structures shown. The upper curve ($\bar{M}=-35.13$) terminates in the
$T_{8,1}$ elastic torus ribbon knot. Both ends of the dot-dashed curve
($\bar{M}=13.67$) terminate in a double covering of the $T_{4,1}$ elastic
torus ribbon knot, whereas the $T_{8,3}$ structure (shown at the minimum
end-to-end distance of the loop) does not quite close, as can be seen by the
fact that the curve from which it originates does not meet the vertical axis.
The other structure shown on the same loop in the line diagram (shown at the
maximum end-to-end distance of the loop on the figure) is related to the
$T_{8,3}$ torus knot, but it is also not closed, and moreover the surface of
the strip is self-intersecting.

As an example of further strip solutions, in figure~\ref{fig:Surface_neq_4}(b)
is shown a solution, with $n=8$ and under a compressive axial force, that is
not located on any of the computed response curves.

\begin{figure}
\centering
\includegraphics[width=0.99\textwidth]{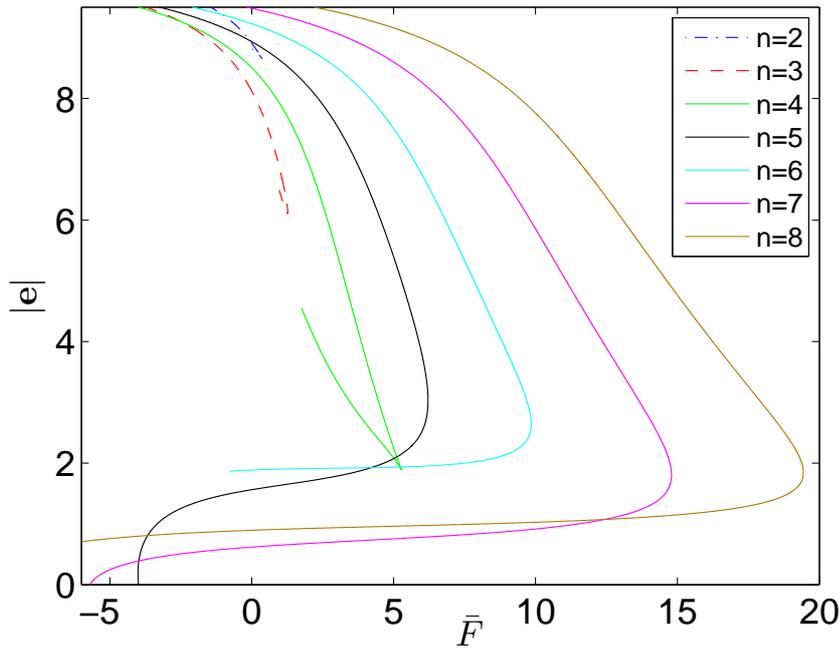}
\caption{
Fold structure of scaled end-to-end distance versus applied force $\bar{F}$, $n=2,\ldots,8$ at constant $\bar{M}=14.49$.
}
\label{fig:bif_n2to8F2}
\end{figure}

Figure~\ref{fig:bif_n2to8F2} displays force-extension curves for varying
mode number, $n=2,\ldots,8$, at fixed moment $\bar{M}$. To obtain this plot,
strips for different modes have been scaled to the same aspect ratio
$2nL/2w=10.5300$. Parts of these curves with positive slope are expected to
correspond to stable solutions. The curves predict that, for $n\geq 5$, under
increasing tension solutions jump to higher mode (down in $|\mathbf{e}|$) as
the force is increased beyond the folds seen in the diagram.

\section{Discussion}
When twisted, and pulled, an acetate model strip buckles into a regular
pattern of triangular facets.
We have computed periodic solutions describing this buckling pattern
by formulating and numerically solving a geometrically-exact boundary-value
problem for the large deformation of a thin, wide, inextensible strip.
We have also obtained response curves of force against end-to-end distance
and twisting moment against end-to-end angle for mode numbers $n=2$, 4 and 8.
Our results predict critical forces and jumps into higher buckling modes that
would be interesting to explore experimentally.

By construction our solutions are periodic, which tends to be what one
observes in experiments. However, non-periodic solutions can be constructed
in a similar way by matching different trapezoidal segments. One would keep
the first symmetry operation (reflection about the normal at the cylindrical
point) but instead of reflecting about the binormal at the inflection point
in the second step one could match the segment to a (suitably rescaled)
trapezoidal segment of different length $L$. The resulting solution would no
longer be symmetric about the inflection points.

In Mansfield (1989) and Ashwell (1962) analyses were performed similar to
ours in that there too the double strain energy integral was reduced by
integrating along the generator. However, the integrand expression was
subsequently simplified by using an approximate moment balance equation that
ignores the non-planarity of the strip. This approach gives a realistic
evolution of the generator with arclength, but fails to predict the
appearance of a cylindrical point at finite distance from the clamped end
of the strip.

Triangular patterns are known to occur in a variety of problems of elastic
sheets, including fabric draping and paper crumpling (Witten 2007). A sheet
crumples (forming sharp points and straight creases) when it is forced into a
constrained area. The sheet is predominantly under compression. It is
interesting to note that, by constrast, in our triangular buckling pattern
the strip is in (relatively high) tension. In both cases we observe a
focussing of the strain energy which may lead to fracture of the material.
Strain energy localisation thus appears to be a generic response of a thin
elastic sheet to an external constraint.

Our results may be relevant for paper, fabric and sheet metal processing.
They may also be of importance in the robotic manipulation of flexible belts,
e.g., film circuit boards (Wakamatsu \textit{et al}. 2007). In all these sheet
manipulations it is important to avoid shapes with high concentrated stresses
that may lead to tearing. Our formulation may help to choose boundary
conditions that avoid unwanted configurations.

\begin{acknowledgements}
This work was supported by the UK's Engineering and Physical Sciences Research
Council (EPSRC) under grant no. EP/F023383/1.
\end{acknowledgements}

\appendix{Derivation of the equations using standard variational techniques}
We wish to calculate the variation of the functional $H=\int f(\kappa,\tau,\kappa',\tau')\, \ud s$. 
This calculation can be done using the formulation of Capovilla \textit{et al}. (2002), substituting explicit expressions for quantities wherever they arise and grouping similar terms, which gives fully simplified expressions for all quantities of interest.

Alternatively, one can re-use the variations already calculated in
Capovilla \textit{et al}. (2002), to yield the same results, without having
to compute new variations. There, for example, the variation is given for
$H_1=\int f_1(\kappa,\tau)\, \ud s$, stating that it can be written down from
the variations of $H_2=\int f_2(\kappa)\, \ud s$ and
$H_3=\int f_3(\tau)\, \ud s$, which are previously calculated. The resulting
force and moment corresponding to $H_1$ can then be written down from the
expressions for the force and moment for $H_2$ and $H_3$. This follows from
the chain rule. Similarly, one can calculate the variation of the functional
$H=\int f(\kappa,\tau,\kappa',\tau')\, \ud s$ using $\delta H_1$ without
computing new variations. This follows as a consequence of the chain and
product rules in differentiation. The resulting force and moment
corresponding to $H$ can then be written down from the expressions for the
force and moment for $H_1$, with a small number of additional terms arising
from the chain and product rules.

Instead of following these approaches, we here use a third and more economical
method by leaving $\delta H$ expressed in terms of $\delta\kappa$ and
$\delta\tau$ only. This method is easily generalised to functionals involving
higher derivatives of $\kappa$ and $\tau$. Following the notation of
Capovilla \textit{et al}. (2002), the infinitesimal deformation of a space
curve is $\delta\mathbf{r}=\Psi_{||}\mathbf{t}+\Psi_1\mathbf{n}+\Psi_2\mathbf{b}$,
and denoting $\delta_{||}, \delta_{\perp}$ as the tangential and normal parts
of the deformation respectively, the variation of any functional
$H=\int f \, \ud s$ is
\begin{equation}\label{var_H1}
\delta H=\delta_0 H+\int \delta_{\perp}f\, \ud s, 
\end{equation}
%
where $\delta_0 H\equiv\delta_{||}H+\int f\delta_{\perp}\, \ud s=\int \left((f\Psi_{||})'-f\kappa\Psi_1\right)\, \ud s$.

We therefore need to calculate the second term on the right-hand side of equation \eqref{var_H1}, where by the chain rule,
$\delta_{\perp}f=f_{\kappa}\delta_{\perp}\kappa+f_{\tau}\delta_{\perp}\tau+f_{\kappa'}\delta_{\perp}\kappa'+f_{\tau'}\delta_{\perp}\tau'$, using the notation $f_{\beta}\equiv\partial f/\partial\beta$. 
Since for any scalar $h$, $\delta_{\perp}(h')=\kappa h'\Psi_1+(\delta_{\perp}h)'$, then
\begin{equation}\label{var_deltaf}
\delta_{\perp}f=(f_{\kappa}-f'_{\kappa'})\delta_{\perp}\kappa+(f_{\tau}-f'_{\tau'})\delta_{\perp}\tau
+\kappa(\kappa'f_{\kappa'}+\tau'f_{\tau'})\Psi_1
+(f_{\kappa'}\delta_{\perp}\kappa+f_{\tau'}\delta_{\perp}\tau)',
\end{equation}
where $\delta_{\perp}\kappa$ and $\delta_{\perp}\tau$ are given in Capovilla \textit{et al}. (2002).
Equation \eqref{var_H1} along with equation \eqref{var_deltaf} is of the form 
\begin{equation}\label{var_H}
\delta H=\int \ud s\, \mathcal{E}_i\Psi_i+\int \ud s\, \mathcal{Q}',
\end{equation}
with $\mathcal{Q}$ the Noether charge, so that the Euler-Lagrange equations $\mathcal{E}_i=0$ can be immediately written down, since these are just the coefficients of $\Psi_i$ which can be read off from equation \eqref{var_H1}. These are two coupled differential equations in the unknowns $\kappa,\tau$, and the known density $f$, which are quoted in general form in Thamwattana \textit{et al}. (2008) and in Hangan (2005) for the Sadowsky functional.

Apart from the Euler-Lagrange equations, it is of interest to find
expressions for the conserved force $\bm{F}$ and the moment $\bm{M}$. The
force $\bm{F}$ is obtained by specialising the deformation to a constant
infinitesimal translation $\delta\mathbf{r}=\mathbf{e}$, where $\bm{F}$ is
defined by $\mathcal{Q}=-\mathbf{e}\cdot\bm{F}$. Similarly, the conserved
moment $\bm{T}$ is obtained by specialising the deformation to a constant
infinitesimal rotation $\delta\mathbf{r}=\mathbf{\Omega}\times\mathbf{r}$,
where $\bm{T}$ is defined by $\mathcal{Q}=-\mathbf{\Omega}\cdot\bm{T}$ and
decomposed into $\bm{T}=\mathbf{r}\times\bm{F}+\bm{M}$. Thus only the total
derivative parts of \eqref{var_H1} contribute to $\bm{F}$ and $\bm{M}$.
But these contributions can be read off from the results in
Capovilla \textit{et al}. (2002) by noting that a term in equation
\eqref{var_deltaf} of the form $a\delta_{\perp}b$, where $b=\kappa, \tau$,
has to be re-expressed in the form \eqref{var_H} to isolate its total
derivative part. This has already been done in
Capovilla \textit{et al}. (2002). For a term in equation \eqref{var_deltaf}
of the form $(a\delta_{\perp}b)'$, one just needs the contribution from
$\delta_{\perp}b$.

Thus, for an infinitesimal rotation
$\delta\mathbf{r}'=\mathbf{\Omega}\times\mathbf{r}'=\mathbf{\Omega}\times\mathbf{t}$
and $\delta\mathbf{r}''=\kappa\mathbf{\Omega}\times\mathbf{n}$.
Using the compact expressions of Langer \& Perline (1991) (equations (3.7b,c)),
$\delta\kappa=\delta\mathbf{r}''\cdot\mathbf{n}-2\kappa\delta\mathbf{r}'\cdot\mathbf{t}$
gives $\delta\kappa=0$, i.e.,
$\delta_{\perp}\kappa=-\delta_{||}\kappa=-\Psi_{||}\kappa'$.
But $\Psi_{||}\equiv\delta\mathbf{r}\cdot\mathbf{t}=(\mathbf{r}\times\mathbf{t})\cdot\mathbf{\Omega}$,
giving $\delta_{\perp}\kappa=-\mathbf{\Omega}\cdot(\kappa'\mathbf{r}\times\mathbf{t})$.
As this is of the form $-\mathbf{\Omega}\cdot(\mathbf{r}\times\mathbf{F})$,
$\mathcal{Q}\equiv a\delta_{\perp}\kappa$ in \eqref{var_deltaf} does not
contribute to the moment $\bm{M}$. Similarly,
$\delta\tau=\delta\mathbf{r}''\cdot\mathbf{b}/\kappa'+\delta\mathbf{r}'\cdot(\kappa\mathbf{b}-\tau\mathbf{t})=0$,
giving $\delta_{\perp}\tau=-\mathbf{\Omega}\cdot(\mathbf{r}\times\tau'\mathbf{t})$.
As this is of the form $-\mathbf{\Omega}\cdot(\mathbf{r}\times\mathbf{F})$,
$\mathcal{Q}\equiv a\delta_{\perp}\tau$ in \eqref{var_deltaf} also does not
contribute to the moment $\bm{M}$. (Note that by retaining these terms one
would obtain the force $\mathbf{F}$, but in the next paragraph we will,
instead, calculate $\mathbf{F}$ by considering a constant infinitesimal
translation.)
Now, from Capovilla \textit{et al}. (2002) (equations (49) and (67)),
$a\delta_{\perp}\kappa$ contributes $-a\mathbf{b}$
to the moment, whereas $a\delta_{\perp}\tau$
contributes $-a\mathbf{t}-a'\mathbf{n}/\kappa$. Thus from
equations \eqref{var_H1} and \eqref{var_deltaf} one can immediately write
down the moment as:

\begin{equation}\label{eqs_M}
 \bm{M}=(f'_{\tau'}-f_{\tau})\mathbf{t}+(f''_{\tau'}-f'_{\tau})\mathbf{n}/\kappa+(f'_{\kappa'}-f_{\kappa})\mathbf{b}.
\end{equation}

In order to calculate the contribution to the force of
$\mathcal{Q}\equiv f_{\kappa'}\delta_{\perp}\kappa + f_{\tau'}\delta_{\perp}\tau$, 
one notes that since
$\delta\mathbf{r}=\mathbf{e}$ is a constant, it follows from 
the formulation of Langer \& Perline (1991) (equations (3.7b,c)) that
$\delta\kappa=0$, i.e.,
$\delta_{\perp}\kappa=-\delta_{||}\kappa=-\Psi_{||}\kappa'=-\mathbf{e}\cdot(\kappa'\mathbf{t})$.
Thus the contribution of $\mathcal{Q}\equiv f_{\kappa'}\delta_{\perp}\kappa$ in \eqref{var_deltaf} to $\bm{F}$ is $\kappa' f_{\kappa'}\mathbf{t}$. In a similar manner, for this constant translation, $\delta\tau=0$, giving $\delta_{\perp}\tau=-\mathbf{e}\cdot(\tau'\mathbf{t})$, so that $\mathcal{Q}\equiv f_{\tau'}\delta_{\perp}\tau$ in \eqref{var_deltaf} contributes $\tau' f_{\tau'}\mathbf{t}$ to the force.
Now from Capovilla \textit{et al}. (2002) (equations (48) and (64)),
$a\delta_{\perp}\kappa$ contributes $\kappa a\mathbf{t}+a'\mathbf{n}+\tau a\mathbf{b}$
to the force, whereas $a\delta_{\perp}\tau$ contributes
$\tau a\mathbf{t}+\tau a'\mathbf{n}/\kappa-((a'/\kappa)'+\kappa a)\mathbf{b}$. Also, $\delta_{||} H$ contributes $-f\mathbf{t}$ to the force.
Thus from equations \eqref{var_H1} and \eqref{var_deltaf} one can immediately write down the force as:
\begin{equation}\label{eqs_F}
\begin{split}
\bm{F}& =\left(-f+\kappa' f_{\kappa'}+\tau' f_{\tau'}+\kappa(f_{\kappa}-f'_{\kappa'})+\tau(f_{\tau}-f'_{\tau'})\right)\mathbf{t}\\
      & +\quad \left(f'_{\kappa}-f''_{\kappa'}+\frac{\tau}{\kappa}(f'_{\tau}-f''_{\tau'})\right)\mathbf{n}\\
      & +\quad \left(\tau(f_{\kappa}-f'_{\kappa'})-\kappa(f_{\tau}-f'_{\tau'})-
        ((f'_{\tau}-f''_{\tau'})/\kappa)'\right)\mathbf{b}.
\end{split}
\end{equation}

From these components of $\bm{F}$ and $\bm{M}$, the differential equations
for the moment components and the differential equation for $F_t$ in equation
\eqref{eqs_vector_b} are easily obtained. For instance, from \eqref{eqs_M} it
is seen that $M_t'=\kappa M_n$, while equations for $M_n$, $M_b$ and $F_t$
can be extracted similarly. The remaining two equations for $F_n$ and $F_b$
are obtained from \eqref{eqs_F} and the Euler-Lagrange equations
$\mathcal{E}_i=0$.

In summary, the variation of $H=\int f(\kappa,\tau,\kappa',\tau')\, \ud s$
can be easily obtained from the variations $\delta\kappa, \delta\tau$. Their
explicit expressions are not required; their contributions can be read off
from Capovilla \textit{et al} (2002). The additional terms generated by the
additional dependence of $f$ on $\kappa'$ and $\tau'$ are therefore easily
managed.

One way of regarding equations \eqref{eqs_M} and \eqref{eqs_F}, is that they
prescribe $\bm{F}$ and $\bm{M}$ once $\kappa$ and $\tau$ for the given
density $f$ are known. The two Euler-Lagrange equations for $\kappa$ and
$\tau$ are given by $\mathcal{E}_i=0$, i.e., by setting the coefficients of
$\Psi_i$ to zero in equation \eqref{var_H}. Instead of solving the problem
this way, however, it is preferable to set up an equivalent system of coupled
one-dimensional ODEs as in Starostin \& van~der Heijden (2009). One way to
do this is to use the natural variables $\bm{F}$ and $\bm{M}$, and, for the
functional \eqref{energy5}, to change variables from $(\kappa,\tau)$ to
$(\kappa,\eta)$.

By considering functionals of the form
$g(\kappa,\eta,\eta')=f(\kappa,\tau,\kappa',\tau')$, one can show that
$(\partial g/\partial\kappa)_{\eta,\eta'}=f_{\kappa}+\eta f_{\tau}+\eta' f_{\tau'}$, $(\partial g/\partial\eta)_{\kappa,\eta'}=\kappa f_{\tau}+\kappa' f_{\tau'}$ and $(\partial g/\partial\eta')_{\kappa,\eta}=\kappa f_{\tau'}$.
Using these identities, and writing the components of the internal force
$\bm{F}$ and moment $\bm{M}$ in the directions of the Frenet frame of
tangent, principal normal and binormal as $\bm{F}=(F_t,F_n,F_b)^T$,
$\bm{M}=(M_t,M_n,M_b)^T$, one can show, using the tangent and binormal
components of $\bm{M}$ from equation \eqref{eqs_M}, that
\begin{equation}\label{eqs_scalar0}
\begin{split}
\partial_\kappa g+\eta M_t+M_b&=0,\\
\left(\partial_{\eta'} g\right)'-\partial_\eta g-\kappa M_t&=0.
\end{split}
\end{equation}

The two scalar equations, \eqref{eqs_scalar0}, along with the six balance
equations (equation \eqref{eqs_vector_b}) for the components of the internal
force $\bm{F}$ and moment $\bm{M}$ (see Starostin \& van~der Heijden 2007,
Starostin \& van~der Heijden 2009), are then the equations satisfied by the
centreline of the developable strip for the unknowns
$\bm{F},\bm{M},\kappa,\eta$:
\begin{eqnarray}
&& \hspace{-1cm} \bm{F}'+\bm{\omega}\times\bm{F}=\bm{0}, \qquad
\bm{M}'+\bm{\omega}\times\bm{M}+\bm{t}\times\bm{F}=\bm{0},
\label{eqs_vector} \\
&& \hspace{-1cm} \partial_\kappa g+\eta M_t+M_b=0, \qquad
\left(\partial_{\eta'} g\right)'-\partial_\eta g-\kappa M_t=0,
\label{eqs_scalar}
\end{eqnarray}
where $\bm{\omega}=\kappa(\eta,0,1)^T$ is the Darboux vector. 
Equations \eqref{eqs_vector}, \eqref{eqs_scalar} constitute a system of
differential-algebraic equations that are turned into a system of ODEs by
differentiation of the algebraic equation in \eqref{eqs_scalar}.

\label{lastpage}
\end{document}